\documentclass[fp, twocolumn]{jpsj3}
\usepackage{txfonts}
\usepackage{stfloats}
\usepackage{braket}
\usepackage{color}
\usepackage[normalem]{ulem}

\title{Analysis of Multifrequency Oscillating Magnetic Fields by Neutron Spin Interferometry}

\author{Ryuto Fujitani$^{1,2}$\thanks{fujitani.ryuto.85m@st.kyoto-u.ac.jp}, Masahiro Hino$^2$, and Takashi Higuchi$^{2,3}$}
\inst{$^1$Department of Nuclear Engineering, Kyoto University, Kyoto 615-8246, Japan \\
$^2$Institute for Integrated Radiation and Nuclear Science (KURNS), Kyoto University, Kumatori, Osaka 590-0494, Japan \\
$^3$Research Center for Nuclear Physics (RCNP), The University of Osaka, Osaka 567-0047, Japan }

\abst{
We have developed a methodology for analyzing multifrequency oscillating magnetic fields using neutron spin interferometry.
A theoretical formulation was derived for the contrast and the phase of the interference pattern for an input oscillating magnetic field consisting of a series of sine functions with multiple frequencies.
According to the formulation, the contrast decays because of the dispersion of the Larmor precession induced by an oscillating field.
The formulation also predicts that the phase of the interference pattern is not constant in contrast to single-frequency cases.
We conducted an experiment on the JRR-3 C3-1-2-2 beamline to confirm the formulation in the cases of double-frequency oscillating fields in the frequency range from 2500 to 10000~Hz.
The measured interference patterns showed reasonable agreement with the theoretical predictions.
}

\begin{document}
\maketitle

\section{Introduction} 

Neutrons are electrically neutral and have magnetic moments. 
This gives them both high penetration and sensitivity to magnetic fields, making them effective probes in materials science\cite{NeutronDataBooklet, Kardjilov_2011, utsuro2010handbook}.
Among the methods employing neutrons, neutron spin interferometry (NSI) is one of the most sensitive methods for detecting magnetic fields.
It relies on the detection of the Larmor precession phase between the two spin states as neutrons traverse a sample\cite{baryshevskii1991neutron, badurek1986neutron, frank2002neutron, utsuro2010handbook, YAMAZAKI2002623, Hino_1999, Hino_199903, achiwa2002neutron, utsuro2005neutron, Hasegawa_2003, bartosik2009experimental, shen2020unveiling}. 
Techniques based on this principle, such as neutron spin echo (NSE), are widely used to study dynamics in soft matter, magnetic systems, and other media\cite{mezei2002neutron, golub1987neutron, gahler1992neutron}.

Neutrons are also useful in magnetic field imaging owing to their high penetration ability. 
From the perspective of energy efficiency, measurements of oscillating magnetic fields in materials, such as electric motors and transformers, are becoming important. 
The use of stroboscopic systems for the polarized neutron imaging of periodic, varying magnetic fields has also been studied\cite{Tremsin_2015, Hiroi_2015}. 
However, stroboscopic systems are limited by time scales. 
In previous work, we proposed a method for analyzing single-frequency oscillating magnetic fields in the kilohertz range using neutron spin interferometry with continuous neutron exposure\cite{suzuki-fujitani}. 
In this study, we extend this approach to multifrequency oscillating magnetic fields. We present the analytical formulation and demonstrate the validity of the proposed approach through experiments.

\section{Neutron Spin Interferometry}
In neutron spin interferometry,
the spin states of polarized neutrons are manipulated by resonant spin flippers (RSFs) to create and control the superposition of their two spin states.
When a neutron enters into an interferometer,
it is polarized by a polarizer and transformed into a superposition state by the first RSF.
The state of a neutron, $\psi$, after the first RSF can be written as
\begin{equation}\label{eq:superposition1}
  \psi = \frac{1}{\sqrt{2}}\left(\Ket{+}-e^{i\chi_1}\Ket{-}\right),
\end{equation}
where $\Ket{+}$ and $\Ket{-}$ donate the spin-up and spin-down states, 
respectively, and $\chi_1$ is the phase of the oscillating field generated by the first RSF.
After passing through a magnetic field, 
the second RSF again generates a superposition of the spin states.
Then, an analyzer polarizes a neutron:
\begin{equation}
  \psi = \frac{1}{2}(e^{-i\phi}-e^{i(\phi-\chi_2+\chi_1)})\Ket{+},
\end{equation}
where $\chi_2$ is the phase of the oscillating field generated by the second RSF 
and $\phi$ is the phase induced by the magnetic field between two RSFs.
Note that a neutron experiences a potential of the opposite sign depending on its spin state. 
Consequently, the magnetic field generates a phase between the two spin states.
By defining $\chi = \chi_2 - \chi_1$, the detection probability of a neutron is given by
\begin{equation}
  |\psi|^2 = \frac{1}{2}\left\{1-\cos(\chi-2\phi)\right\}.
\end{equation}
Thus, a sinusoidal interference pattern can be obtained.
The experimentally observed interference pattern is shown in Fig.~\ref{fig:testFringe}.

In experimental measurements, this pattern is observed as the neutron count $N$, 
which can be expressed as 
\begin{equation}\label{eq:NSI_pattern}
  N = \frac{N_0}{2} \left\{ 1 - \cos (\chi - \Omega - \delta) \right\},
\end{equation}
where $N_0$ is the neutron count without RSFs.
The terms $\delta$ and $\Omega$ represent the phases induced by the guide magnetic field and the magnetic field between RSFs,
respectively.
These phases satisfy $2\phi = \delta + \Omega$.
Since a guide magnetic field is applied throughout the neutron spin interferometer system to maintain spin orientation,
a consistent phase exists across measurements.
To distinguish the guide field contribution, the constant phase offset $\delta$ is introduced.
The phase shift $\Omega$ is expressed by the following integral of the magnetic field between RSFs:
\begin{equation}\label{eq:MF_Integral}
  \Omega = \frac{2|\mu_n|}{\hbar v} \int^{l}_{0} B(x) dx,
\end{equation}
where $\hbar$ is Dirac's constant, $\mu_n$ is the magnetic moment of the neutron, 
$v$ is the neutron velocity, and $l$ is the distance between RSFs.
$B(x)$ is the strength of the magnetic field along the neutron path $x$ between the RSFs.

In practice, owing to the finite polarization efficiencies of the polarizer and analyzer, 
as well as the spin-flip efficiency of RSFs, 
the contrast $C$ of an interference pattern does not reach unity. 
Therefore, the experimentally observed interference pattern is described by
\begin{equation}\label{eq:NSI_pattern_fit}
  N = \frac{N_0}{2} \left\{ 1 - C \cos (\chi - P) \right\},
\end{equation}
where the phase of an interference pattern is $P$. 
Parameters $C$, $P$ and $N_0$ can be extracted by fitting the measured neutron count to Eq.(\ref{eq:NSI_pattern_fit}).

\section{Theoritical Description}
\subsection{Formulation of multifrequency oscillating magnetic field}
A multifrequency oscillating magnetic field $B(x,t)$ at position $x$ and time $t$ can be expressed as a Fourier series expansion based on the fundamental frequency $f$.
\begin{equation}\label{eq:magneticField}
  B(x,t) = B_0(x)\sum_{n=1}^{\infty} d_n \sin(-2\pi nft + \phi_n)
\end{equation}
Here, $\phi_n$ denotes the phase of the $n$-th frequency component oscillation 
and $d_n$ is the amplitude of the $n$-th component normalized by the fundamental amplitude ($d_1 = 1$).
$B_0(x)$ represents the strength of the magnetic field without time dependence.

When a neutron enters a region of an oscillating magnetic field at time $\tau$, 
in the same way as in Eq.~(\ref{eq:MF_Integral}), the phase shift $\Omega$ observed in spin interference is given by
\begin{equation}\label{eq:OmegaTheta1}
  \Omega(\tau) = \frac{2|\mu_n|}{\hbar v} \int_{0}^{l} B_0(x)\sum_{n=1}^{\infty} d_n \sin\left(-2\pi fn (t - \tau) + \phi_n\right) dx.
\end{equation}
Since the magnetic field oscillates while a neutron traverses the field,
the magnetic field experienced by a neutron is obtained by substituting $t = x/v$ into Eq.~(\ref{eq:OmegaTheta1}).
Thus, Eq.~(\ref{eq:OmegaTheta1}) can be rewritten as
\color{black}
\begin{equation}\label{eq:OmegaTheta}
  \Omega(\theta) = \frac{2|\mu_n|}{\hbar v} \int_{-\infty}^{\infty} B_0(x)\sum_{n=1}^{\infty} d_n \sin\left(-\frac{2\pi f}{v} nx + n\theta + \phi_n\right) dx.
\end{equation}
Here, $\theta$ represents the phase of the oscillating magnetic field at the moment a neutron enters the field, defined by $\theta = 2\pi f \tau$.
By assuming $B_0(x) = 0$ at $x \leq 0$ or $x \geq l$, the integration range has been extended to infinity.
Equation~(\ref{eq:OmegaTheta}) provides the expression of $\Omega$ for an oscillating magnetic field,
corresponding to the static case in Eq.~(\ref{eq:MF_Integral}).
Introducing
\begin{equation}\label{eq:k}
  k = 2\pi f /v
\end{equation}
and using the $k$-space Fourier transform of the magnetic field profile $B_{0}(x)$, 
the power spectrum $|b(k)|$ and the phase $\theta(k)$ are defined as
\begin{equation}\label{eq:bk}
  |b(k)|e^{i\theta(k)} = \int_{-\infty}^{\infty} B_0(x) e^{-ikx} dx.
\end{equation}
Extracting only the imaginary part of Eq. \eqref{eq:bk} gives
\begin{equation}\label{eq:bk_sin}
  |b(k)|\sin(\theta(k)) = \int_{-\infty}^{\infty} B_0(x) \sin(-kx) dx.
\end{equation}
Thus, the phase shift $\Omega(\theta)$ can be rewritten as
\begin{equation}\label{eq:Omega}
  \Omega(\theta) = \frac{2|\mu_n|}{\hbar v} \sum_{n=1}^{\infty} |b(nk)| d_n \sin\left(n\theta + \phi_n + \theta(nk)\right).
\end{equation}

\subsection{Formulation of interference pattern}
Consider an interference pattern obtained for an oscillating magnetic field with continuous exposure.
In the case of continuous exposure, the phase $\theta$ can take any value within the range $0 \leq \theta \leq 2\pi$.
Therefore, the observed interference pattern corresponds to an average over all the values of $\theta$.
Consequently, an observed interference pattern is given by
\begin{equation}
  N = \frac{N_0}{2} \frac{1}{2\pi} \int_{0}^{2\pi} \left\{ 1 - \cos (\chi - \Omega(\theta) - \delta) \right\} d\theta.
\end{equation}
This equation describes an integration of the dispersion of the phase shift.
Introducing the complex value $c$ defined as
\begin{equation}\label{eq:intExpOmega}
  c = \frac{1}{2\pi} \int_{0}^{2\pi} \exp{(i\Omega(\theta))} d\theta,
\end{equation}
an interference pattern can be writen as
\begin{equation}\label{eq:NSI_c}
  N = \frac{N_0}{2} \left\{ 1 - \left| c \right| \cos (\chi - \delta - \arg\left( c \right)) \right\}.
\end{equation}
The value of $c$ is
\begin{equation}\label{eq:solvedc}
  \begin{aligned}
    c = &\sum_{\substack{m_1 + 2m_2 + 3m_3 + \cdots = 0\\ m_k \in \mathbb{Z}}} J_{m_1}\left(\frac{2|\mu_n|}{\hbar v}|b(k)|\right) J_{m_2}\left(\frac{2|\mu_n|}{\hbar v}|b(2k)|d_2\right)\cdots \\
    &\times \exp\left\{i(m_1 (\phi_1 + \theta(k)) + m_2 (\phi_2 + \theta(2k)) + \cdots)\right\},
  \end{aligned}
\end{equation}
where $\mathbb{Z}$ denotes the set of all integers and $J_{m}(x)$ denotes the $m$-th order Bessel function of the first kind.

The derivation of Eq.~(\ref{eq:solvedc}) is as follows. 
We use the generating function of the Bessel function, given by
\begin{equation}
  \exp \left(\frac{u}{2}\left(w-\frac{1}{w}\right)\right) = \sum_{m=-\infty}^{\infty} J_m(u) w^m,
\end{equation}
where $w$ and $u$ are arbitrary complex numbers, with $w \neq 0$.
By multiplying both sides, we obtain
\begin{equation}\label{eq:besselGenMulti}
  \prod_{n=1}^{\infty} \exp \left(\frac{u_n}{2}\left(w_n-\frac{1}{w_n}\right)\right) = \prod_{n=1}^{\infty} \sum_{m_n=-\infty}^{\infty} J_{m_n}(u_n) w_n^{m_n}.
\end{equation}
Setting $w_n = e^{i(n\theta+\phi_n+\theta(nk))}$ and $u_n = \frac{2|\mu_n|}{\hbar v}|b(nk)|d_n$, the left-hand side of Eq.~(\ref{eq:besselGenMulti}) becomes
\begin{equation}
  \exp\left(i\sum^{\infty}_{n=1} \frac{2|\mu_n|}{\hbar v}|b(nk)|d_n \sin(n\theta+\phi_n+\theta(nk)) \right) = \exp\left(i\Omega (\theta)\right), 
\end{equation}
while the right-hand side of Eq.~(\ref{eq:besselGenMulti}) becomes
\begin{equation}
  \prod_{n=1}^{\infty} \sum_{m_n=-\infty}^{\infty} J_{m_n}\left(\frac{2|\mu_n|}{\hbar v}|b(nk)|d_n\right) \exp\left( im_n(n\theta+\phi_n+\theta(nk)) \right).
\end{equation}
Integrating both sides with respect to $\theta$ from $0$ to $2\pi$, 
and making use of the orthogonality in $\theta$,
we obtain Eq.~(\ref{eq:solvedc}).

\subsection{Case of single-frequency oscillation}
Consider the case in which a single-frequency oscillation is applied.
Under this condition, $d_n = 0$ for $n \geq 2$.
Using the properties of the Bessel function,
 $J_0(0) = 1$ and $J_n(0) = 0 $ for $n \neq 0$, 
Eq.(\ref{eq:solvedc}) becomes
\begin{equation}\label{eq:arbit_1x}
  c  =  J_{0}\left(\frac{2|\mu_n|}{\hbar v} |b(k)|\right).
\end{equation}
This result corresponds to the expression previously reported\cite{suzuki-fujitani}. 
Since $c$ is always real and $\arg\left( c \right)$ shown in Eq.~(\ref{eq:NSI_c}) is always zero,
the phase of the interference pattern remains constant, 
while only the contrast decays following the zero-th order Bessel function.

\subsection{Case of fundamental and second-harmonic oscillations}
Consider fundamental and second-harmonic oscillations given by
\begin{equation}
  B(x, t) = B_0(x) \left\{\sin(-2\pi ft+\phi_1) + d_2 \sin(-4\pi ft+\phi_2)\right\}.
\end{equation}
Introducing $d_n = 0$ for $n \geq 3$,
Eq.~(\ref{eq:solvedc}) becomes
\begin{equation}\label{eq:arbit_2x}
  \begin{aligned}
    c & = \sum_{m = -\infty}^{\infty} J_{-2m}\left(\frac{2|\mu_n|}{\hbar v} |b(k)|\right) J_{m}\left(\frac{2|\mu_n|}{\hbar v} |b(2k)|d_2\right) \\
    &\times \exp\left\{i(-2m(\phi_1 + \theta(k)) + m(\phi_2 + \theta(2k)))\right\}.
  \end{aligned}
\end{equation}
Since $c$ can be a complex value and $\arg\left( c \right)$ shown in Eq.~(\ref{eq:NSI_c}) is not always zero, the phase of the interference pattern can be shifted because of the oscillating magnetic field.

\section{Experimental Setup}
To demonstrate Eqs.~(\ref{eq:arbit_1x}) and (\ref{eq:arbit_2x}), 
we conducted an experiment on the JRR-3 C3-1-2-2 (Multilayer Interferometer and reflectometer for NEutron 2, MINE2) beamline.
The incident neutron beam is continuous and monochromated to a wavelength of 8.8~$\AA$ and the full width at half maximum of $2.7\%$.

\begin{figure}
  \begin{center}
  \includegraphics[width=\linewidth]{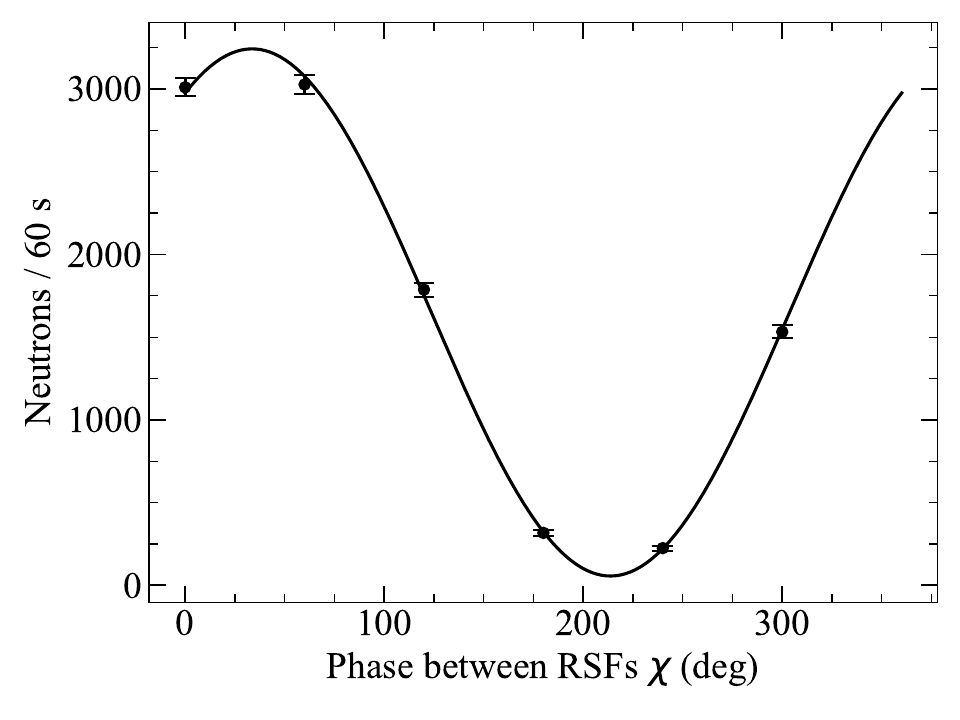}
  \caption{Interference pattern obtained from the spin interferometer shown in Fig.~\ref{fig:interferometerSetup}. }\label{fig:testFringe}
  \end{center}
\end{figure}

\begin{figure*}[t]
 \begin{center}
  \includegraphics[width=\textwidth]{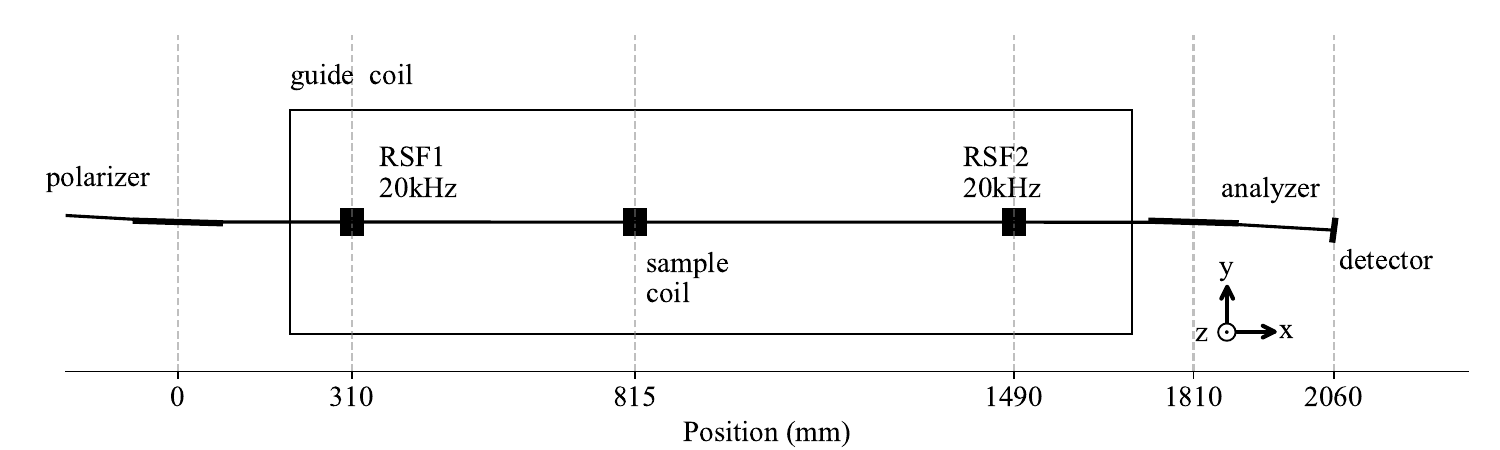}
  \caption{Experimental setup for a neutron spin interferometer at JRR-3 C3-1-2-2. }
  \label{fig:interferometerSetup}
 \end{center}
\end{figure*}

The experimental setup is shown in Fig.~\ref{fig:interferometerSetup}.
The beam direction is along the $x$-axis.
The setup of the spin interferometer consists, in order from the upstream, of a polarizer, RSF1, a sample coil, RSF2, an Sanalyzer, and a neutron detector.
A guide magnetic field is applied throughout the entire spin interferometer system to maintain the neutron spin quantization axis.

\begin{figure}
  \begin{center}
  \includegraphics[width=\linewidth]{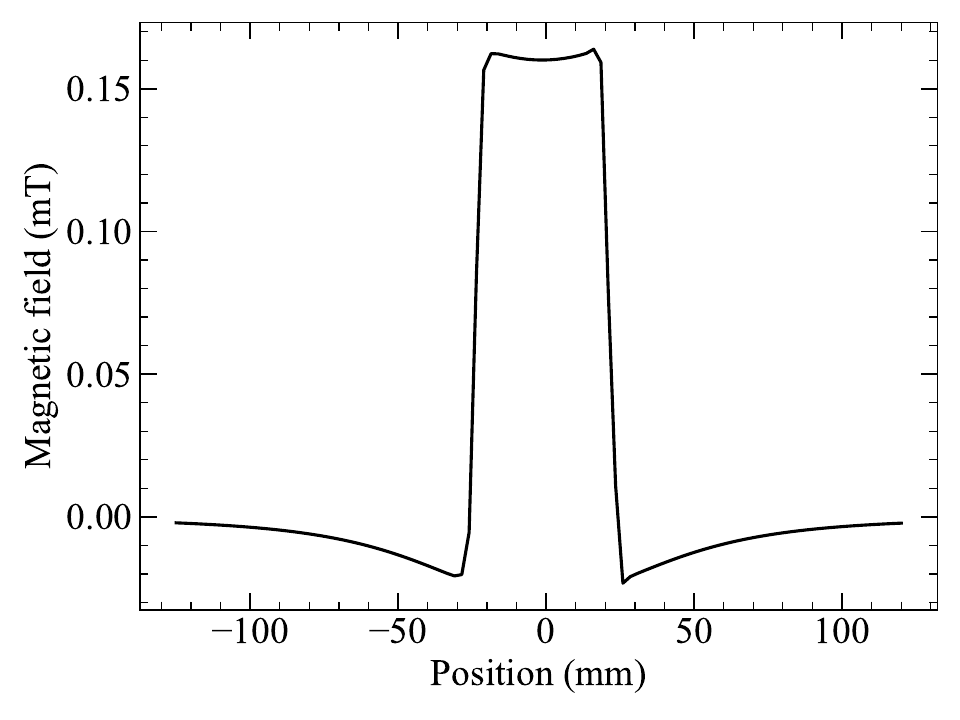}
  \caption{Magnetic field generated by sample coil when 1~A current is applied. The calculation was performed using the Biot--Savert law.}\label{fig:Bz}
  \end{center}
\end{figure}

The polarizer and analyzer, which reflect only spin-up neutrons, are composed of magnetic multilayer mirrors.
The RSFs, which flip the neutron spin orientation, generate 20~kHz oscillating magnetic fields along the $x$-axis.
The oscillating magnetic field to be analyzed by the spin interferometry is produced by a rectangle coil named the sample coil.
Its dimensions are 100~mm in the $z$-direction (height), 
45~mm in the $x$-direction (length), and 70~mm in the $y$-direction (width), and it consists of 15 turns.
Its wire is 1~mm thick and is made of Al, enabling lower neutron absorption.
The spatial distribution of the magnetic field at the center of the coil along the $x$-axis calculated by using the Biot--Savert law for a current of 1~A is shown in Fig.~\ref{fig:Bz}.

An interference pattern obtained with this setup is shown in Fig.~\ref{fig:testFringe}.
Fitting the data to Eq.~(\ref{eq:NSI_pattern_fit}), 
the contrast $C$ of the interference pattern is determined to be $0.9659 \pm 0.0077$ and the phase $P$ is $33.65 \pm 0.64$~deg.

To demonstrate Eq.~(\ref{eq:arbit_1x}),
we applied a single-frequency oscillating current to the coil. 
Interference patterns were measured while varying the amplitude of the AC current from 0.0 to 3.0~A peak-to-peak ($\rm{A_{pp}}$). 
Measurements were carried out at frequencies from 2500 to 20000~Hz.

Subsequently, to demonstrate Eq.~(\ref{eq:arbit_2x}),
we applied an AC current composed of fundamental and second-harmonic oscillations.
We measured interference patterns while varying the phase $\phi_2$ of a double-frequency oscillation shown in Eq.~(\ref{eq:arbit_2x}) from 0 to 360~deg.  
The amplitude of AC current for a fundamental oscillation was fixed to 2~$\rm{A_{pp}}$, 
while the relative amplitude of a second-harmonic oscillation, denoted as $d_2$, was varied among 0.25, 0.5, 0.75 and 1. 
Measurements were performed for fundamental frequencies from 2500, 5000, 7500 and 10000~Hz.

\section{Results and Discussion}

\subsection{Case of single-frequency oscillation}

\begin{figure}
  \begin{center}
  \includegraphics[width=\linewidth]{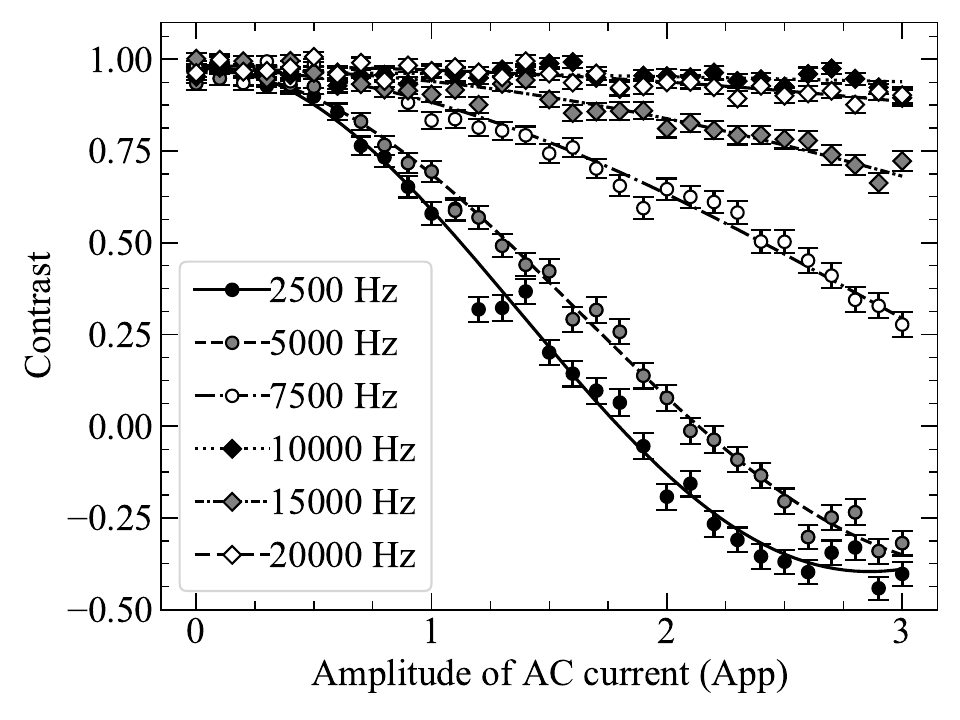}
  \caption{Contrast of interference patterns varying with the amplitude of the sinusoidal AC current applied to the coil.}\label{fig:bessel}
  \end{center}
\end{figure}

In Fig.~\ref{fig:bessel}, the contrast follows the zero-th order Bessel function of the applied current as predicted by Eq.~(\ref{eq:arbit_1x}).
The vertical axis in Fig.~\ref{fig:bessel} represents the contrast of the interference pattern, 
while the horizontal axis indicates the amplitude of the AC current applied to the coil.
Data for each frequency are plotted on the same graph.
The plots represent experimental data, which have been scaled by a factor of $1/0.9659$,
the inverse of the contrast observed without the sample, as shown in Fig.~\ref{fig:testFringe}.
By introducing the power spectrum $|b_{\rm{1A_{pp}}}(k)|$ for a current of 1~$\rm{A_{pp}}$ and the AC current amplitude $I$, Eq.~(\ref{eq:arbit_1x}) can be rewritten as
\begin{equation}\label{eq:arbit_1x_rewrite}
  J_{0} \left(\frac{2|\mu_n|}{\hbar v}|b_{\rm{1A_{pp}}}(k)|  \times I \right).
\end{equation}
The lines represent fits to Eq.~(\ref{eq:arbit_1x_rewrite}).
By fitting data to Eq.~(\ref{eq:arbit_1x_rewrite}), the value of $2|\mu_n| |b_{\rm{1A_{pp}}}(k)| / \hbar v$ can be extracted.

\begin{figure}
  \begin{center}
  \includegraphics[width=\linewidth]{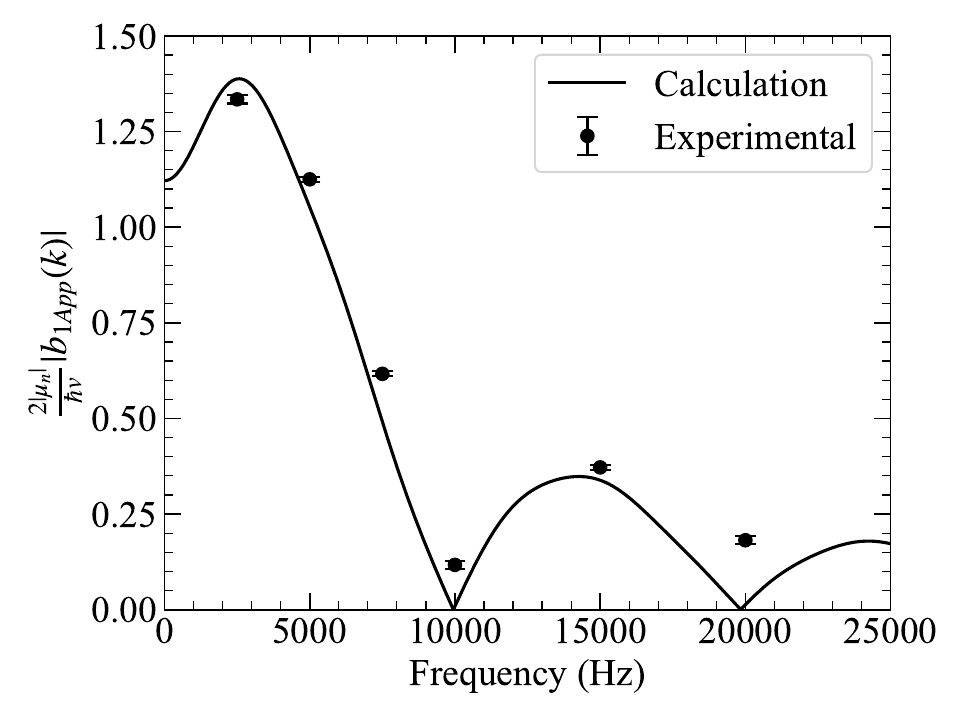}
  \caption{Calculated and measured values of $2|\mu_n||b_{\rm{1A_{pp}}}(k)| / \hbar v$. 
  The calculated values are obtained from $B(k)$ shown in Fig.~\ref{fig:Bz} using the Biot--Savart law and a neutron velocity corresponding to 8.8~\AA. These are compared with the measured values shown in Fig.~\ref{fig:bessel}.}\label{fig:bk_compare}
  \end{center}
\end{figure}

Figure~\ref{fig:bk_compare} shows the values of $2|\mu_n| |b_{\rm{1A_{pp}}}(k)| / \hbar v$ as a function of the frequency of the oscillating magnetic field.
The line represents the value calculated using the magnetic field shown in Fig.~\ref{fig:Bz}.
The plots are obtained from the measured results shown in Fig.~\ref{fig:bessel}.
Since $k$ is proportional to the frequency, 
the values of $2|\mu_n| |b_{\rm{1A_{pp}}}(k)| / \hbar v$ have a frequency dependence.

The experimental data reproduce the model well.
The remaining discrepancy can be considered to be imperfections in the coil shape. 
While the calculated values were obtained using the Biot--Savart law, 
distortions in the coil or its installation angle can affect the value of $|b_{\rm{1A_{pp}}}(k)|$.

\subsection{Case of fundumental and second-harmonic oscillations}

\begin{figure}
  \begin{center}
  \includegraphics[width=\linewidth]{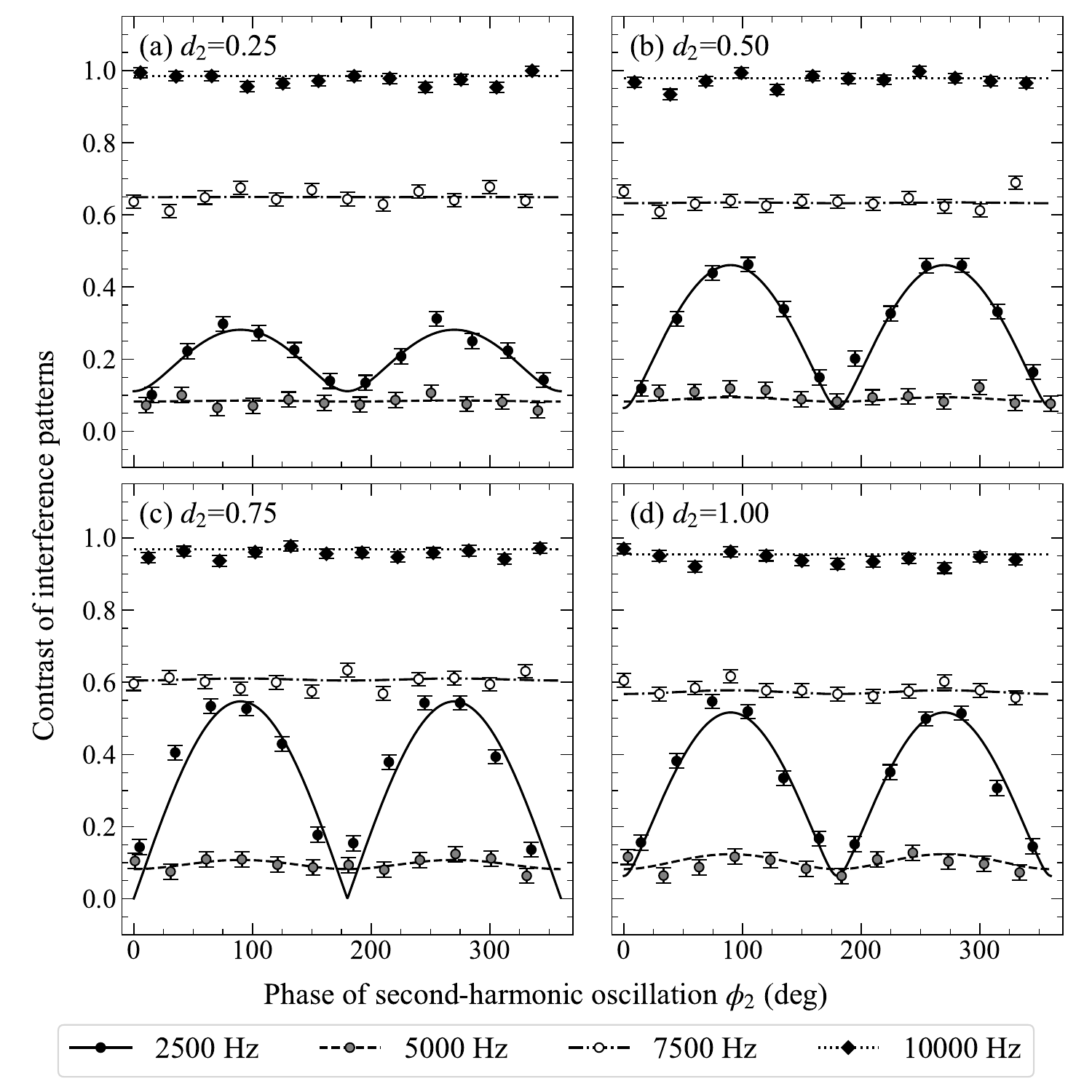}
  \caption{Contrast of interference patterns when the amplitude of a fundamental oscillation is 2~$\rm{A_{pp}}$, where (a) $d_2$ = 0.25, (b) $d_2$ = 0.5, (c) $d_2$ = 0.75, and (d) $d_2$ = 1. To calculate the estimation lines, the measured $2|\mu_n| |b_{\rm{1A_{pp}}}(k)| / \hbar v$ values shown in Fig.~\ref{fig:bk_compare} were used. }\label{fig:phi2ConVsD_2App}
  \end{center}
\end{figure}

\begin{figure}
  \begin{center}
  \includegraphics[width=\linewidth]{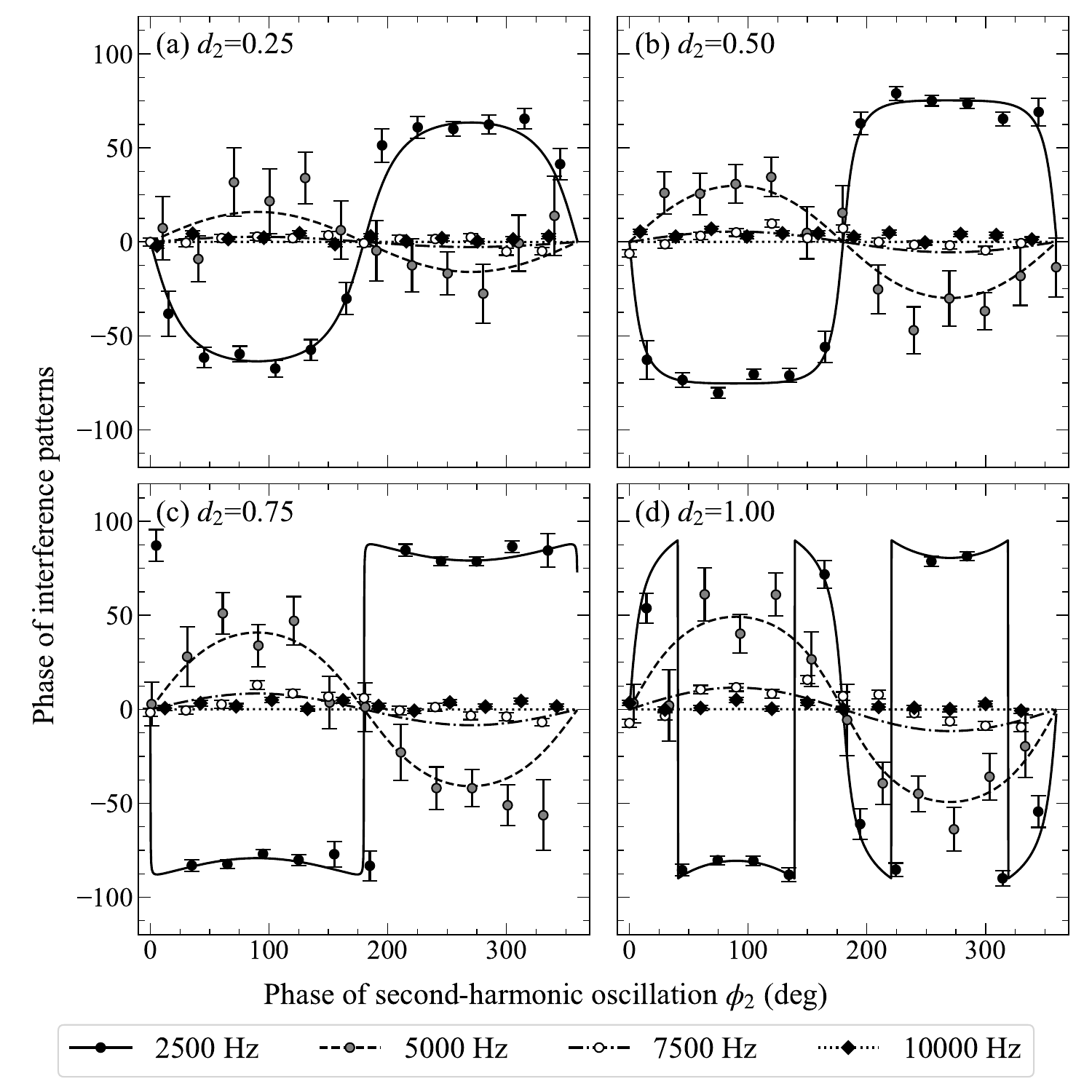}
  \caption{Phase of interference patterns when the amplitude of a fundamental oscillation is 2~$\rm{A_{pp}}$, where (a) $d_2$ = 0.25, (b) $d_2$ = 0.5, (c) $d_2$ = 0.75, and (d) $d_2$ = 1. To calculate the estimation lines, the measured $2|\mu_n| |b_{\rm{1A_{pp}}}(k)| / \hbar v$ values shown in Fig.~\ref{fig:bk_compare} were used.}\label{fig:phi2PhaVsD_2App}
  \end{center}
\end{figure}

The contrast measured as a function of the phase of a second-harmonic oscillation, $\phi_2$, is shown in Fig.~\ref{fig:phi2ConVsD_2App}, 
and the phase is shown in Fig.~\ref{fig:phi2PhaVsD_2App}.
These results correspond to the case where the fundamental oscillation of the applied current is 2~$\rm{A_{pp}}$.
Here, the frequency in the legends refers to that of a fundamental oscillation.
The contrast is scaled by a factor of $1/0.9659$.
The lines represent the results calculated using Eq.~(\ref{eq:arbit_2x}) and the measured $2|\mu_n| |b_{\rm{1A_{pp}}}(k)| / \hbar v$ in Fig.~\ref{fig:bk_compare}.
In the calculation, the summation was performed over the range $m = -20$ -- $20$.

In the experiments, the AC current was generated by the function generator and a power amplifier.
Because of the inductance, determining the phase of a second-harmonic oscillation, $\phi_2$, is difficult.
Therefore, in Figs.~\ref{fig:phi2ConVsD_2App} and \ref{fig:phi2PhaVsD_2App},
the experimental plot is shifted horizontally to the calculation results with phase $\phi_2$.

Both the contrast and phase exhibit periodic behavior with respect to $\phi_2$, but they are not analytically simple such as sinusoidal. 
Despite the complexity of these variations, the experimental results show reasonable reproducibility.

For data with the fundamental frequency of 2500~Hz, the contrast and phase shown in Figs.~\ref{fig:phi2ConVsD_2App} and \ref{fig:phi2PhaVsD_2App} exhibit the most significant changes among the four frequencies.
Changes in the phase of a second-harmonic oscillation depend on the value of $2|\mu_n| |b_{\rm{1A_{pp}}}(2k)| / \hbar v$ for the corresponding second harmonic.
As shown in Fig.~\ref{fig:bk_compare}, the value of $2|\mu_n| |b_{\rm{1A_{pp}}}(k)| / \hbar v$ at 5000~Hz is the highest among those at 5000, 10000, 15000, and 20000~Hz, which are the second-harmonic frequencies of 2500, 5000, 7500, and 10000~Hz.
Consequently, the data for 2500~Hz show the most significant changes.
In contrast, the values of $2|\mu_n| |b_{\rm{1A_{pp}}}(k)| / \hbar v$ are smaller at 10000, 15000, and 20000~Hz.
Therefore, the data for fundamental frequencies of 5000, 7500, and 10000~Hz show slight changes in  contrast and phase.

To evaluate the effective frequency range, 
we consider a rectangular magnetic field $B_0 (x)$ whose length is $L$, frequency is $f$, and strength is $B$:
\begin{equation}
  B_0 (x) = \begin{cases} 
    B & 0 \le x \le L \\ 
    0 & \text{otherwise}.
    \end{cases}
\end{equation}
From Eq.~(\ref{eq:bk}), the following equation can be obtained:
\begin{equation}\label{eq:bk_rect}
  \frac{2|\mu_n|}{\hbar v} |b(k)| = \frac{2|\mu_n|}{\hbar} \frac{B}{\pi f L} \left| \sin \left(\frac{\pi f L}{v} \right) \right|.
\end{equation}
Equation~(\ref{eq:bk_rect}) shows that $2|\mu_n| |b(2k)| / \hbar v$ vanishes at $fL/v = 1$ and generally decreases as $f$ increases.
Thus, the fundamental frequency $f$ should satisfy $f < v/L$ in order to obtain measurable changes.
The $n$-th harmonic also remains observable as long as $nf < v/L$.
Although the magnetic field in this experiment is not perfectly rectangular, as shown in Fig.~\ref{fig:Bz}, 
a rough estimation is sufficient for this study.
Given the conditions of $L = 45~\mathrm{mm}$ and $v=450~\mathrm{m/s}$, 
$2|\mu_n||b(k)| / \hbar v $ becomes zero at $f = 10000~\mathrm{Hz}$ and smaller for $f > 10000~\mathrm{Hz}$.
Therefore, the data in Figs.~\ref{fig:phi2ConVsD_2App} and \ref{fig:phi2PhaVsD_2App}, which have fundamental frequencies of 5000, 7500, and 10000~Hz, have small variations. 
Notably, at 10000~Hz, the contrast remains almost unity.

\section{Conclusions}
We presented an approach for analyzing multifrequency sinusoidal oscillating magnetic fields in neutron spin interferometry.
We derived a formulation of the resulting interference patterns assuming the oscillating field to be a multifrequency oscillation represented as a Fourier series expansion.
The contrast and phase of the interference pattern were expressed as sums of sine functions weighted by Bessel functions.
This was confirmed in an experiment with an oscillating magnetic field composed of fundamental and second-harmonic oscillations in the frequency range from 2500 to 10000~Hz.
Although the resulting interference patterns are not analytically simple, 
the experimental data showed reasonable agreement with theoretical predictions, demonstrating the validity of the proposed approach.

\section*{Acknowledgment}
This neutron experiment at JRR-3 was carried out through the JRR-3 general user program managed by the Institute for Solid State Physics, the University of Tokyo (proposal Nos. 24583, 24407).
The fabrication of polarizing devices was conducted under the visiting researcher's program of the Institute for Integrated Radiation and Nuclear Science, Kyoto University.
The development of neutron mirrors was also supported by the visiting researcher’s program of the Research Reactor Institute, Kyoto University and JSPS KAKENHI (Grant No. 23K23274).
This work is also financially supported by the JST FOREST Program (Grant No. JPMJFR2237).

\bibliographystyle{jpsj}
\bibliography{bibliography}

\end{document}